\documentclass[english,aps,manuscript,showpacs,preprint,superscriptaddress]{revtex4-1}
\usepackage{amsmath}
\usepackage{amssymb}
\usepackage{graphicx}

\makeatletter

\usepackage{amsfonts}\usepackage{amsthm}\usepackage{latexsym}
\usepackage{color}
\usepackage{xcolor}\usepackage{subfigure}\usepackage{manfnt}
\usepackage{picinpar}
\usepackage{subeqnarray}
\usepackage{cases}
\usepackage{extarrows}
\usepackage{epstopdf}
\usepackage{picins}

\usepackage{color}
\usepackage{amsmath}
\usepackage{amssymb}
\usepackage{amsthm}

\makeatletter

\newcommand{\rom}[1]{\uppercase\expandafter{\romannumeral#1}}

\newcommand{\sumli}{\sum\limits}

\newcommand{\beq}{\begin{equation}}
\newcommand{\eeq}{\end{equation}}
\newcommand{\bal}{\begin{align}}
\newcommand{\eal}{\end{align}}
\newcommand{\baln}{\begin{align*}}
\newcommand{\ealn}{\end{align*}}

\theoremstyle{plain}\newtheorem{theorem}{Theorem}[section]
\theoremstyle{definition}
\theoremstyle{plain}\newtheorem{lemma}[theorem]{Lemma}

\begin{document}
\title{Hamiltonian particle-in-cell methods for Vlasov-Maxwell equations}
\author{Yang He }

\affiliation{School of Nuclear Science and Technology and Department of Modern Physics, University of Science and
Technology of China, Hefei, Anhui 230026, China}

\affiliation{Key Laboratory of Geospace Environment, CAS, Hefei, Anhui 230026,
CHINA}

\author{Yajuan Sun}

\affiliation{LSEC, Academy of Mathematics and Systems Science, Chinese Academy
of Sciences, P.~O.~Box 2719, Beijing 100190, CHINA}

\author{Hong Qin}
\affiliation{School of Nuclear Science and Technology and Department of Modern Physics, University of Science and
Technology of China, Hefei, Anhui 230026, China}
\affiliation{Plasma Physics Laboratory, Princeton University, Princeton, New Jersey
08543, USA}

\author{Jian Liu}

\affiliation{School of Nuclear Science and Technology and Department of Modern Physics, University of Science and
Technology of China, Hefei, Anhui 230026, China}

\affiliation{Key Laboratory of Geospace Environment, CAS, Hefei, Anhui 230026,
CHINA}

\begin{abstract}
In this paper, we develop Hamiltonian  particle-in-cell methods for Vlasov-Maxwell equations by applying conforming finite element methods in space and splitting methods in time. For the spatial discretisation, the criteria for choosing finite element spaces are presented such that the semi-discrete system possesses a discrete non-canonical Poisson structure.  We apply a Hamiltonian splitting method to the semi-discrete system in time, then the resulting algorithm is  Poisson preserving and explicit. The conservative  properties of the algorithm guarantee the efficient and accurate numerical simulation of the Vlasov-Maxwell equations over long-time.
\end{abstract}

\maketitle
\section{Introduction}

In modern plasma physics and accelerator physics, numerical simulation of the Vlasov-Maxwell (VM) equations
is an  indispensable tool for the study of the interactive dynamics of charged particles with electromagnetic fields. The system of  dimensionless Vlasov-Maxwell equations regardless of the relativistic
effects  reads
\begin{subequations}
\begin{gather}
\frac{\partial f}{\partial t}+\mathbf{v}\cdot\frac{\partial f}{\partial\mathbf{x}}+\left(\mathbf{E}+\mathbf{v}\times\mathbf{B}\right)\cdot\frac{\partial f}{\partial\mathbf{v}}=0,\label{eq:vlasov}\\
\nabla\times\mathbf{B}=\int_{\Omega_{v}}f\mathbf{v}d\mathbf{v}+\frac{\partial\mathbf{E}}{\partial t},\label{eq:MaxwellE}\\
\nabla\times\mathbf{E}=-\frac{\partial\mathbf{B}}{\partial t},\label{eq:MaxwellB}\\
\nabla\cdot\mathbf{E}=\int_{\Omega_{v}}fd\mathbf{v},\label{eq:divE}\\
\nabla\cdot\mathbf{B}=0,\label{eq:divB}
\end{gather}
\end{subequations}
where $f(\mathbf{x},\mathbf{v},t)$ is the distribution function of
position $\mathbf{x}\in\Omega_{x}\subset\mathbb{R}^{3}$ and velocity
$\mathbf{v}\in\Omega_{v}\subset\mathbb{R}^{3}$ at time $t$, and $(\mathbf{E}(\mathbf{x},t),\mathbf{B}(\mathbf{x},t))\in\mathbb{R}^{3}\times\mathbb{R}^{3}$
are the electromagnetic fields. 
As the distribution function $f$ is posed in a 6 dimensional phase space, the numerical computation is quite consuming and challenging for high dimensional problems. The presentation of particle-in-cell (PIC)  methods \cite{Hockney88, Birdsall91} greatly reduce the computation amount. In the PIC method, the Vlasov equation is solved by following particles' trajectories on  Lagrangian grids, and the fields are approximated on  Eulerian grids. It has been widely applied for decades. In most cases, the most concerned characteristics of the VM equations is the long term behaviours and multi-scale structures, then it is crucial to improve the stability and reliability of the PIC methods over long time.
Conventionally, the Vlasov equation (\ref{eq:vlasov}) and Maxwell's equations (\ref{eq:MaxwellE}-\ref{eq:MaxwellB}) are solved by standard numerical methods, such as 4-th order Runge-Kutta methods. The local energy-momentum conservation laws can not be preserved, and after long time of  computation the accumulation of the error leads to non-reliable results. To overcome this difficulty, we apply geometric integration methods to the coupled equations.

Geometric integration methods \cite{Ruth83,Feng85,Feng10,Hairer03} are designed to conserve the intrinsic properties inherited  by the original systems, including the Hamiltonian (symplectic and Poisson) structure, invariant phase space volume, etc. It has been confirmed   that  such methods show superior behavior for long term  simulation \cite{Hairer03}. Specifically, explicit symplectic and volume-preserving algorithms are developed for solving  single particle motions \cite{Qin08-PRL,Qin09-PoP,chin08-PRE,Qin13-084503,He15-JCP,He16-JCP,Zhang14,Zhang15-pop,He15KSym}, numerical experiments exhibit that they can bound the error of energy and simulate the trajectory of the particle well over long time. For the Vlasov-Maxwell equations, geometric methods can be derived based on their different formulations. From the Lagrangian formulation based on the variational principle, variational symplectic algorithms of the VM system are derived \cite{Squire12,Evstatiev13,Kraus14Phd,Shadwick14,Xiao13,Xiao15Pop}. On the other hand, the VM system is an infinite dimensional Hamiltonian system characterized by a Poisson bracket and a Hamiltonian functional \cite{Morrison80,Marsden82}.  In Ref.~\onlinecite{Qin16SymVM}, canonical symplectic methods are developed by discretising the canonical Poisson bracket directly. And a discrete non-canonical Poisson bracket is preserved \cite{Xiao15Pop} when the VM system is discretised by the method of discrete exterior calculus in space and a Hamiltonian splitting method in time \cite{He15Pop}.

In this paper, we  further study Hamiltonian methods for the VM equations by combining the PIC technique with finite element methods (FEM). After approximating the distribution function $f$ by Klimontovich representation, we discretise Maxwell's equations (\ref{eq:MaxwellB}) and (\ref{eq:MaxwellE})  by  finite element methods. The choice of FEM serves the purpose of deriving a semi-discrete system equipped with a discrete structure. 
We will show that if the curl space of finite element approximation for the electric field is a subset of the space for the magnetic field, and the approximate magnetic field is divergence free,  the semi-discrete system  is Hamiltonian.  Finite element methods such as the edge element method of N\'{e}d\'{e}lec \cite{Nedelec80}, or the elements from finite element exterior calculus \cite{Arnold14}, etc. can be applied to satisfy this sufficient condition for the existence of a discrete structure. The discrete structure is characterized by a discrete Poisson bracket consistent with the non-canonical Morrison-Marsden-Weinstein bracket \cite{Marsden82}. The general expression of the discrete Poisson bracket and the discrete Hamiltonian is also given.
For the semi-discrete system, we split the system into five parts, and fully discretise the system by combining the exact solutions of the subsystems. This technique is the Hamiltonian splitting method presented by us in Ref.~\onlinecite{He15Pop}. It is proved that the  resulting numerical methods conserve the divergence-free property of the magnetic field, and are Hamiltonian with the discrete Poisson bracket preserved. Furthermore, the update mapping at each time step is explicit, and can be generalized to higher order easily.  

 The outline of this  work is as follows.
In the next section,  a general  formulation of the semi-discrete VM system is presented. The Vlasov equation is
discretised to a particle system in use of the Klimontovich representation, and Maxwell's
equations are discretised in the framework of conforming finite element methods.
In section 3, we present the discrete bracket and the discrete Hamiltonian of the semi-discrete system, and establish a general criteria for the elements to guarantee the bracket Poisson. In section 4, the
Poisson-preserving temporal discretisation of the VM equations is given by using the Hamiltonian splitting.

\section{Spatial discretisation for the Vlasov-Maxwell equations}
We start this section from the discretisation of the distribution function $f$.
The discrete distribution function denoted by $f_D$ is written in a sum of Dirac masses,
\begin{equation}
f_{D}(\mathbf{x},\mathbf{v},t)= \sum_{s}f_{s}=\sum_{s}\omega_{s}\delta(\mathbf{x}-\mathbf{X}_{s})\delta(\mathbf{v}-\mathbf{V}_{s}),\label{eq:Disf}
\end{equation}
where $(\mathbf{X}_{s},\mathbf{V}_{s})$ is  the $s$-th particle's coordinate
in phase space.
Substituting  Eq.~(\ref{eq:Disf}) into  the Vlasov-Maxwell equations (\ref{eq:vlasov})-(\ref{eq:MaxwellB}) leads to the particle-Maxwell equations,
\begin{subequations}
\begin{gather}
\dot{\mathbf{X}}_{s}=\mathbf{V}_{s},\qquad\dot{\mathbf{V}}_{s}=\int_{\Omega_{x}}(\mathbf{E}(\mathbf{x},t)
+\mathbf{V}_{s}\times\mathbf{B}(\mathbf{x},t))\delta(\mathbf{x}-\mathbf{X}_{s})d\mathbf{x},\label{eq:Vlasov-1}\\
\nabla\times\boldsymbol{B}=\sum_{s}\mathbf{V}_{s}\delta(\mathbf{x}-\mathbf{X}_{s})+\frac{\partial\boldsymbol{E}}{\partial t},\label{eq:MaxwellE-1}\\
\nabla\times\boldsymbol{E}=-\frac{\partial\boldsymbol{B}}{\partial t}.\label{eq:MaxwellB-1}
\end{gather}
\end{subequations}
In  practical computation, the Dirac delta function $\delta(\mathbf{x}-\mathbf{X}_{s})$ in Eqs.~(\ref{eq:Vlasov-1}) and (\ref{eq:MaxwellE-1}) is usually   replaced by  a smooth function $S(\mathbf{x}-\mathbf{X}_{s})$ with property $\int_{\Omega_{x}}S(\mathbf{x}-\mathbf{X}_{s})d\mathbf{x}=1$. This can help to reduce the noise from numerical computation.

For the particle-Maxwell equations (\ref{eq:Vlasov-1})-(\ref{eq:MaxwellB-1}), we consider the problem with 
perfect conducting boundary conditions (PEC),
$$\mathbf{n}\times \mathbf{E}=0 \text{~~on~} \partial \Omega_x, \quad \mathbf{ n}\cdot \mathbf{B} =0 \text{~~on~} \partial \Omega_x,$$
where $\mathbf{n}$ is the unit normal vector of the boundary $\partial \Omega_x$ pointing out of the domain $\Omega_x$. The initial conditions at time $t=0$ are
\begin{equation}\label{eq:initial}
\begin{gathered}
\mathbf{E}(\mathbf{x},0)=\mathbf{E}_0(\mathbf{x}), \quad \mathbf{B}(\mathbf{x},0)=\mathbf{B}_0(\mathbf{x}),\\
\text{with~} \nabla\cdot \mathbf{E}_0(\mathbf{x})=\sum_s\mathbf{V}_{s}(0)\delta(\mathbf{x}-\mathbf{X}_{s}(0)), \quad\nabla\cdot \mathbf{B}_0(\mathbf{x})=0.
 \end{gathered}
 \end{equation}
 Considering the properties of the solution and the boundary condition, we will solve field variables $\mathbf{E, B}$ in the function spaces,
\begin{gather*}
\mathcal{E}=\{\mathbf{E}\in H(\mathbf{curl},\Omega_{x}):\mathbf{E}\times \mathbf{n}|_{\partial \Omega_x}=0\},\\
\mathcal{B}=\{\mathbf{B}\in H(\mathbf{div},\Omega_{x}):\mathbf{B}\cdot \mathbf{n}|_{\partial \Omega_x}=0\},
\end{gather*}
where
$H(\mathbf{curl},\Omega_{x})\equiv\{\mathbf{G}\in L^{2}(\Omega_{x})^{3}:\nabla\times\mathbf{G}\in L^{2}(\Omega_{x})^{3}\}$, and $
H(\mathbf{div},\Omega_{x})\equiv\{\mathbf{G}\in L^{2}(\Omega_{x})^{3}:\nabla\cdot\mathbf{G}\in L^{2}(\Omega_{x})\}$.
Multiplying Eq.~(\ref{eq:MaxwellE-1}) and Eq.~(\ref{eq:MaxwellB-1}) by test functions $\Phi\in \mathcal{E}$ and $\Psi\in \mathcal{B}$ respectively, and integrating over $\Omega_x$, we  obtain the variational problem for the continuous Maxwell's equations,
\begin{equation}
\begin{gathered}
\left(\partial_{t}\mathbf{E},\mathbf{\Phi}\right)= \left(\mathbf{B},\nabla\times\mathbf{\Phi}\right)- \left(\sum_{s}\mathbf{V}_{s}\delta(\mathbf{x}-\mathbf{X}_{s}),\mathbf{\Phi}\right),
 \quad \forall \mathbf{\Phi}\in \mathcal{E},\\
\left(\partial_{t}\mathbf{B},\mathbf{\Psi}\right)=-\left(\nabla\times\mathbf{E},\mathbf{\Psi}\right),
\quad \forall \mathbf{\Psi}\in \mathcal{B}.
\end{gathered}
\label{eq:VariForm}
\end{equation}
Here $\left(\mathbf{f},\mathbf{g}\right)=\int_{\Omega_{x}}\mathbf{f}\cdot\mathbf{g}d\mathbf{x}$ denotes the inner product of two vector functions $\mathbf{f}$ and $\mathbf{g}$.  

Next we discretise the problem in Eq.(\ref{eq:VariForm}) by conforming finite element methods. We firstly present a general matrix formulation of the discrete system, then choose the elements in the next section. Let $\mathcal{T}_{h}=\{K\}$ be regular partitions
of the spatial domain $\Omega_{x}$ with  $K$ the Cartesian elements.
Suppose that the field variables  are approximated by
the piecewise polynomials $\mathbf{E}_{h}$ and $\mathbf{B}_{h}$ in the finite element spaces $\mathcal{E}_h$ and $\mathcal{B}_h$ respectively. The approximate problem is then to find $(\mathbf{E}_h, \mathbf{B}_h)\in \mathcal{E}_h\times\mathcal{B}_h$ such that
\begin{equation}
\begin{gathered}
\left(\partial_{t}\mathbf{E}_h,\mathbf{\Phi}\right)= \left(\mathbf{B}_h,\nabla\times\mathbf{\Phi}\right)- \left(\sum_{s}\mathbf{V}_{s}\delta(\mathbf{x}-\mathbf{X}_{s}),\mathbf{\Phi}\right), \quad\forall \mathbf{\Phi}\in\mathcal{E}_h,\\
\left(\partial_{t}\mathbf{B}_h,\mathbf{\Psi}\right)=-\left(\nabla\times\mathbf{E}_h,\mathbf{\Psi}\right),\quad \forall \mathbf{\Psi}\in \mathcal{B}_h.
\end{gathered}
\label{eq:DisVariForm}
\end{equation}
Denoting $\{\mathbf{W}^{e}(\mathbf{x})\}_{j=1}^{N_e}$ the basis functions of $\mathcal{E}_h$ and $\{\mathbf{W}^{b}(\mathbf{x})\}_{j=1}^{N_b}$
the basis functions of  $\mathcal{B}_h$, then
\begin{equation}
\mathbf{E}_{h}(\mathbf{x},t)=\sum_{j=1}^{N_{e}}E_{j}(t)\mathbf{W}_{j}^{e}(\mathbf{x}), \quad\mathbf{B}_{h}(\mathbf{x},t)=\sum_{j=1}^{N_{b}}B_{j}(t)\mathbf{W}_{j}^{b}(\mathbf{x}).\label{eq:DisEB}
\end{equation}
Substituting  Eq.~(\ref{eq:DisEB}) into  Eq.~(\ref{eq:DisVariForm}) and taking  $\mathbf{\Phi}=\mathbf{W}_{i}^{e}$,
$\mathbf{\Psi}=\mathbf{W}_{i}^{b}$ for every $i=1...N_{e}(N_{b})$ gives the discretisation of Maxwell's equations,
\begin{equation}
\begin{aligned} & \sum_{j=1}^{N_{e}}\left(\mathbf{W}_{j}^{e},\mathbf{W}_{i}^{e}\right)\partial_{t}E_{j}=\sum_{j=1}^{N_{b}}\left(\mathbf{W}_{j}^{b},\nabla\times\mathbf{W}_{i}^{e}\right) B_{j}-\sum_{s}\left(\mathbf{V}_{s}\delta(\mathbf{x}-\mathbf{X}_{s}),\mathbf{W}_{i}^{e}\right),i=1\ldots N_{e}\\
 & \sum_{j=1}^{N_{b}}\left(\mathbf{W}_{j}^{b},\mathbf{W}_{i}^{b}\right)\partial_{t}B_{j}=-\sum_{j=1}^{N_{e}}\left(\nabla\times\mathbf{W}_{j}^{e},\mathbf{W}_{i}^{b}\right) E_{j},i=1\ldots N_{b}.
\end{aligned}
\label{eq:semiDis}
\end{equation}
Similarly, by substituting  Eq.~(\ref{eq:DisEB}) into  Eq.~(\ref{eq:Vlasov-1}),
we get the discrete particle equation,
\begin{equation}
\begin{aligned} & \dot{\mathbf{X}}_{s}=\mathbf{V}_{s},\\
 & \dot{\mathbf{V}}_{s}=\sum_{j}E_{j}\int_{\Omega_{x}}\mathbf{W}_{j}^{e}\delta(\mathbf{x}-\mathbf{X}_{s})d\mathbf{x} +\mathbf{V}_{s}\times\sum_{j}B_{j}\int_{\Omega_{x}}\mathbf{W}_{j}^{b}\delta(\mathbf{x}-\mathbf{X}_{s})d\mathbf{x}.
\end{aligned}
\label{eq:semiDisII}
\end{equation}
We rewrite Eq.~(\ref{eq:semiDis}) and Eq.~(\ref{eq:semiDisII}) as, 
\begin{equation}
\begin{aligned}{\dot{\mathbf{X}}}_{s} & =\mathbf{V}_{s},\\
\dot{\mathbf{V}}_{s} & =\mathcal{W}_{e}^{S}(\mathbf{X}_{s})\mathbf{E}_{D}+\mathbf{V}_{s}\times(\mathcal{W}_{b}^{S}(\mathbf{X}_{s})\mathbf{B}_{D}),\\
\dot{\mathbf{E}}_{D} & =\mathcal{W}_{e}^{-1}\left(\mathcal{K}\mathbf{B}_{D}-\sum_{s}\omega_{s}\mathcal{W}_{e}^{S}(\mathbf{X}_{s})\mathbf{V}_{s}\right),\\
\dot{\mathbf{B}}_{D} & =-\mathcal{W}_{b}^{-1}\mathcal{K}^{\mathrm{T}}\mathbf{E}_{D},
\end{aligned}
\label{eq:semiVM}
\end{equation}
where $\mathbf{E}_{D}:=[E_{1},E_{2},\ldots,E_{N_{e}}]^{\mathrm{T}}$
and $\mathbf{B}_{D}:=[B_{1},B_{2},\ldots,B_{N_{b}}]^{\mathrm{T}}$
denote the  values of the approximate fields. The other matrices are defined as follows:
\begin{itemize}\label{item:Matrices}
\item $\mathcal{W}_{e}$ is an $N_{e}\times N_{e}$ constant symmetric matrix
with $(\mathcal{W}_{e})_{ij}=\left(\mathbf{W}_{i}^{e},\mathbf{W}_{j}^{e}\right)$;
\item $\mathcal{W}_{b}$ is an $N_{b}\times N_{b}$ constant symmetric matrix
with $(\mathcal{W}_{b})_{ij}=\left(\mathbf{W}_{i}^{b},\mathbf{W}_{j}^{b}\right)$;
\item $\mathcal{W}_{e}^{S}(\mathbf{X}_{s})$ is a $N_{e}\times3$ matrix
function with the $j$-th row being $\int_{\Omega_{x}}\left(\mathbf{W}_{j}^{e}\right)^{\top}\delta(\mathbf{x}-\mathbf{X}_{s})d\mathbf{x}$;
\item $\mathcal{K}$ is an $N_{e}\times N_{b}$ constant matrix with $(\mathcal{K})_{ij}=\left(\nabla\times\mathbf{W}_{i}^{e},\mathbf{W}_{j}^{b}\right)$.
\end{itemize}
The discrete form of equations (\ref{eq:divE})-(\ref{eq:divB}) can also be given. 

In the discretisation process, the setting of using conforming finite element methods requires that the spaces have $\mathcal{E}_h\subset \mathcal{E}$, and $\mathcal{B}_h\subset\mathcal{B}$. 
In addition, we need to choose the element in order that  the semi-discrete system   can possess a discrete Poisson structure. This is analyzed in the next section.
\section{Poisson structure of the semidiscrete VM system}

It is known that the  continuous VM equations  can be written in the Hamiltonian formulation by the Morrison-Marsden-Weinstein (MMW) bracket \cite{Marsden82},
\begin{equation*}
\frac{\partial\mathcal{F}}{\partial t}=\{\{\mathcal{F},\mathcal{H}\}\},\label{eq:PoissonVM}
\end{equation*}
with $\mathcal{F}$ being any functional defined on $\mathcal{MV}=\left\{ (f, {\bf E},{\bf B})|\nabla\cdot{\bf B}=0,\nabla\cdot{\bf E}=\int fd^{3}{\bf v}\right\} $,
and $\mathcal{H}$ being the global energy functional. The MMW bracket is Poisson and is preserved by the exact solution of the VM system. In this section,  we establish the conditions which should be satisfied by the finite element basis aiming for the semi-discrete system to have the corresponding discrete Poisson structure.

Similar to the continuous system, we can also write the semi-discrete system (\ref{eq:semiVM}) in
the  form
\[
\dot{F}=\{F,H\}(\mathbf{X}_{s},\mathbf{V}_{s},\mathbf{E}_{D},\mathbf{B}_{D}),
\]
where $F$ is any smooth function of the discrete variables. In this case,  $\{\cdot,\cdot\}$ is the discrete bracket operator defined by
\begin{equation}
\begin{aligned}
&\left\{ F,G\right\} (\mathbf{X}_{s},\mathbf{V}_{s},\mathbf{E}_{D},\mathbf{B}_{D})\\
 & =\sum_{s}\frac{1}{\omega_{s}}\left(\frac{\partial F}{\partial\mathbf{X}_{s}}\cdot\frac{\partial G}{\partial\mathbf{V}_{s}}-\frac{\partial G}{\partial\mathbf{X}_{s}}\cdot\frac{\partial F}{\partial\mathbf{V}_{s}}\right)\\
 & +\left(\frac{\partial F}{\partial\mathbf{E}_{D}}\right)^{\mathrm{T}}\mathcal{W}_{e}^{-1}\mathcal{K}\mathcal{W}_{b}^{-1}\frac{\partial G}{\partial\mathbf{B}_{D}}-\left(\frac{\partial G}{\partial\mathbf{E}_{D}}\right)^{\mathrm{T}}\mathcal{W}_{e}^{-1}\mathcal{K}\mathcal{W}_{b}^{-1}\frac{\partial F}{\partial\mathbf{B}_{D}}\\
 & +\sum_{s}\left(\left(\frac{\partial G}{\partial\mathbf{E}_{D}}\right)^{\mathrm{T}}\mathcal{W}_{e}^{-1}\mathcal{W}_{e}^{S}(\mathbf{X}_{s})\frac{\partial F}{\partial\mathbf{V}_{s}}-\left(\frac{\partial F}{\partial\mathbf{E}_{D}}\right)^{\mathrm{T}}\mathcal{W}_{e}^{-1}\mathcal{W}_{e}^{S}(\mathbf{X}_{s})\frac{\partial G}{\partial\mathbf{V}_{s}}\right)\\
 & +\sum_{s}\frac{1}{\omega_{s}}\mathbf{B}_{D}^{T}\mathcal{W}_{b}^{S}(\mathbf{X}_{s})\left(\frac{\partial F}{\partial\mathbf{V}_{s}}\times\frac{\partial G}{\partial\mathbf{V}_{s}}\right),
\end{aligned}
\label{eq:dPoisson}
\end{equation}
and $H$ is the discrete Hamiltonian,
\begin{equation}\label{eq:ham}
H(\mathbf{X}_{s},\mathbf{V}_{s},\mathbf{E}_{D},\mathbf{B}_{D}) =\frac{1}{2}\sum_{s}\omega_{s}\mathbf{V}_{s}^{2} +\frac{1}{2}\mathbf{E}_{D}^{T}\mathcal{W}_{e}\mathbf{E}_{D}+\frac{1}{2}\mathbf{B}_{D}^{T}\mathcal{W}_{b}\mathbf{B}_{D}.
\end{equation}
It is easy to verify that the discrete bracket in Eq.~(\ref{eq:dPoisson}) and Hamiltonian in Eq.~(\ref{eq:ham}) are consistent with the continuous ones in Ref.~\onlinecite{Marsden82}, and the discrete energy $H$ is an invariant of the semi-discrete system.  The detailed derivation of the discrete bracket is shown in Appendix~\ref{app:1}.

Moreover, if the discrete bracket (\ref{eq:dPoisson}) is skew-symmetric and satisfies the Jacobi identity, it is Poisson and defines a Poisson structure\cite{Hairer03}. This establishes a condition for the basis of the finite element spaces. We have the following lemma.
\begin{lemma}\label{th:Poisson}
The discrete bracket  defined in Eq.~(\ref{eq:dPoisson}) is Poisson if the basis functions $\mathbf{W}_i^e$, $\mathbf{W}_j^b$ of the element spaces for $\mathbf{E}$ and $\mathbf{B}$ satisfy
\begin{subequations}
\begin{gather}
\int\left(\nabla\times\mathbf{W}^e_{i}-\sum_{j=1}^{N_b}\left(\mathcal{K}\mathcal{W}_{b}^{-1}\right)_{ij}\mathbf{W}^b_{j} \right)\delta  ({\bf x}-{\bf X}_s) d\mathbf{x}=0,\quad\forall i,s\label{eq:PoissCondI}\\
\int\nabla\cdot\left(\sum_{j=1}^{N_{b}}B_{j}\mathbf{W}^b_{j}\right)\delta({\bf x}-{\bf X}_s) d{\bf x}=0,\label{eq:PoissCondII}
\end{gather}
\end{subequations}
where $(\mathcal{W}_{b})_{ij}=\left(\mathbf{W}_{i}^{b},\mathbf{W}_{j}^{b}\right)$ and $(\mathcal{K})_{ij}=\left(\nabla\times\mathbf{W}_{i}^{e},\mathbf{W}_{j}^{b}\right)$ are matrices.
\end{lemma}
The proof of the lemma is in Appendix~\ref{app:2}.

In the above lemma, it is observed that  conditions (\ref{eq:PoissCondI}) and (\ref{eq:PoissCondII}) are consistent with  Eqs.~(\ref{eq:MaxwellB}) and (\ref{eq:divB}) respectively.
Notice that if $\nabla\times W_i^e$ can be  expressed as a linear combination of the basis $W_j^b$, i.e. if  $\nabla\times\mathcal{E}_h\subset \mathcal{B}_h$, the condition (\ref{eq:PoissCondI}) holds  naturally. 
We have the following results  for the elements.

\begin{theorem}\label{coro1}
If the space of elements $\mathcal{E}_h$, $\mathcal{B}_h$ for the field variables $\mathbf{E}$ and $\mathbf{B}$  satisfy
\begin{subequations}
\begin{gather}
\nabla\times \mathcal{E}_h\subset \mathcal{B}_h,\label{eq:PoissCond-1}\\
\nabla\cdot\mathbf{B}_h={0},\label{eq:PoissCond-2}
\end{gather}\end{subequations}
then the semi-discrete system (\ref{eq:semiVM}) is a Hamiltonian system with the Poisson bracket  defined in Eq.~(\ref{eq:dPoisson}).
\end{theorem}
In fact, for any fixed point $\mathbf {x}\in \Omega_x$, the discrete equation $\dot{\mathbf{B}}_{D}=-\mathcal{W}_{b}^{-1}\mathcal{K}^{\mathrm{T}}\mathbf{E}_{D}$ in Eq.~(\ref{eq:semiVM}), together with the relation in Eq.~(\ref{eq:PoissCondI}) and Eq.~(\ref{eq:DisEB}) implies that
\[
\frac{d}{dt}\mathbf{B}_h({\bf x})
=\sum_{j=1}^{N_{b}}\dot{B}_{j}\mathbf{W}^b_{j}({\bf x})
=-\nabla\times \mathbf{E}_h({\bf x}).
\]
It follows $\nabla\cdot\mathbf{B}_h(\mathbf{x},t)=\nabla\cdot\mathbf{B}_h(\mathbf{x},0)=0$ along the exact time evolution of $B_j$.
{By analyzing the conditions listed in Theorem \ref{coro1}, we  conclude that the semi-discrete system (\ref{eq:semiVM})  can conserve   the Poisson structure if the finite element spaces  satisfy Eq.~(\ref{eq:PoissCond-1}) and the divergence of the initial magnetic field vanish.
We now list some choices of elements that satisfy Eq.~(\ref{eq:PoissCond-1})  for cubical meshes $\mathcal{T}_h=\{K\}$:
\begin{itemize}
\item Raviart--Thomas--N\'{e}d\'{e}lec's mixed elements\cite{Nedelec80}.
The finite element spaces are
\begin{gather*}
\mathcal{E}_h=\{\mathbf{E}_h\in H(\mathrm{curl,\Omega_x}),\mathbf{E}_h|_K\in Q_{k-1,k,k}\times Q_{k,k-1,k}\times Q_{k,k,k-1}\},\\
\mathcal{B}_h=\{\mathbf{B}_h\in H(\mathrm{div,\Omega_x}),\mathbf{B}_h|_K \in Q_{k,k-1,k-1}\times Q_{k-1,k,k-1}\times Q_{k-1,k-1,k}\},
\end{gather*}
where $Q_{l,m,n}$ denotes the space of polynomials in position variables $(x_1,x_2,x_3)$, with the maximum degree being $l$ for $x_1$, $m$ for $x_2$ and $n$ for $x_3$.
The degrees of freedom are for $K$, and  each edge and face of $K$ (see Ref.~\onlinecite{Nedelec80} for more detail).
\item The elements from finite element exterior calculus. The spaces are
\begin{gather*}
\mathcal{E}_h=S_{k+1}\Lambda^1=\{\mathbf{E}_h|_K=\mathbf{u}+(x_2x_3(w_2-w_3),x_3x_1(w_3-w_1),x_1x_2(w_1-w_2))+\nabla s\},\\
\mathcal{B}_h=S_{k}\Lambda^2=\{\mathbf{B}_h|_K=\mathbf{v}+\nabla\times(x_2x_3(w_2-w_3),x_3x_1(w_3-w_1),x_1x_2(w_1-w_2))\},
\end{gather*}
where $\mathbf{v}_i\in P_k(K)$, $\mathbf{u}_i\in P_{k+1}(K)$, and ${w}_i\in P_k(K)$ independent of $\mathbf{x}$. Here $P_k(K)$ denotes the space of polynomials with degree no higher than $k$. The degrees of freedom can refer to Ref.~\onlinecite{Arnold14}.
\end{itemize}}

\section{Temporal discretisation for the VM equations}
With the appropriate finite element method presented in the above section,  we can derive a semidiscretised system (\ref{eq:semiVM})  with a non-canonical Poisson bracket. It should be  stressed  here that traditional time integrations generally cannot be  applied directly to systems  with  non-canonical Poisson bracket in the purpose of preserving  the structure. However, via  investigating the discrete Hamiltonian in Eq.~(\ref{eq:ham}) it is noticed that our concerned system can be decomposed as a summation of solvable parts. This helps us  to construct the Poisson-preserving methods by  Hamiltonian splitting method presented  in \cite{He15Pop}.

Firstly, we split the Hamiltonian in Eq.~(\ref{eq:ham}) as five parts,
\begin{align*}
 H=H_{E}+H_{B}+H_{1}+H_{2}+H_{3},
\end{align*}
where  $H_{E}=\frac{1}{2}\mathbf{E}_{D}^{T}\mathcal{W}_{e}\mathbf{E}_{D},$
 $H_{B}=\frac{1}{2}\mathbf{B}_{D}^{T}\mathcal{W}_{b}\mathbf{B}_{D}$ and
 $H_{i}=\frac{1}{2}\sum_{s}V_{s}[i]^{2}$.
Here, $V[i]$ denotes the $i$-th Cartesian component of the velocity $\mathbf{V}$. Each part of the Hamiltonian associates with a solvable subsystem.
The subsystem associated with the Hamiltonian $H_{E}$  is $\dot{F}=\left\{ F,H_{E}\right\} $. It is equivalent to
\begin{equation}
\begin{aligned} & \mathbf{\dot{X}}_{s}=0,\\
 & \dot{\mathbf{V}}_{s}=\mathcal{W}_{e}^{S}(\mathbf{X}_{s})\mathbf{E}_{D},\\
 & \dot{\mathbf{E}}_{D}=0,\\
 & \dot{\mathbf{B}}_{D}=-\mathcal{W}_{b}^{-1}\mathcal{K}^{\mathrm{T}}\mathbf{E}_{D}.
\end{aligned}
\label{eq:dSysHE}
\end{equation}
The exact update mapping of this subsystem  with step size
$\Delta t$ is
\begin{equation}
\phi^{E}(\Delta t):\begin{aligned} & \mathbf{X}_{s}(t+\Delta t)=\mathbf{X}_{s}(t),\\
 & \mathbf{V}_{s}(t+\Delta t)=\mathbf{V}_{s}(t)+\Delta t\left(\mathcal{W}_{e}^{S}(\mathbf{X}_{s})\mathbf{E}_{D}\right),\\
 & \mathbf{E}_{D}(t+\Delta t)=\mathbf{E}_{D}(t),\\
 & \mathbf{B}_{D}(t+\Delta t)=\mathbf{B}_{D}(t)-\Delta t\left(\mathcal{W}_{b}^{-1}\mathcal{K}^{\mathrm{T}}\mathbf{E}_{D}(t)\right).
\end{aligned}
\label{eq:dSolHE}
\end{equation}
The equation $\dot{F}=\left\{ F,H_{B}\right\} $ associated with the
Hamiltonian $H_{B}$ is equivalent to
\begin{equation}
\begin{aligned} & \mathbf{\dot{X}}_{s}=0,\\
 & \dot{\mathbf{V}}_{s}=0,\\
 & \dot{\mathbf{E}}_{D}=\mathcal{W}_{e}^{-1}\mathcal{K}\mathbf{B}_{D},\\
 & \dot{\mathbf{B}}_{D}=0.
\end{aligned}
\label{eq:dSysHB}
\end{equation}
The exact update of this subsystem  with step size
$\Delta t$ reads
\begin{equation}
\phi^{B}(\Delta t):\begin{aligned} & \mathbf{X}_{s}(t+\Delta t)=\mathbf{X}_{s}(t),\\
 & \mathbf{V}_{s}(t+\Delta t)=\mathbf{V}_{s}(t),\\
 & \mathbf{E}_{D}(t+\Delta t)=\mathbf{E}_{D}(t)+\Delta t\left(\mathcal{W}_{e}^{-1}\mathcal{K}\mathbf{B}_{D}(t)\right),\\
 & \mathbf{B}_{D}(t+\Delta t)=\mathbf{B}_{D}(t).
\end{aligned}
\label{eq:dSolHB}
\end{equation}
For each $i=1,2,3$, the equation $\dot{F}=\left\{ F,H_{i}\right\} $ associated with
the Hamiltonian $H_i$ is equivalent to
\begin{equation}
\begin{aligned} & \mathbf{\dot{X}}_{s}=V_{s}[i]\mathbf{e}_{i},\\
 & \dot{\mathbf{V}}_{s}=V_{s}[i]\mathbf{e}_{i}\times\left(\mathcal{W}_{b}^{S}(\mathbf{X}_{s})\mathbf{B}_{D}\right),\\
 & \dot{\mathbf{E}}_{D}=\mathcal{W}_{e}^{-1}\left(\sum_{s}\omega_{s}V_{s}[i]\mathcal{W}_{e}^{S}(\mathbf{X}_{s})\mathbf{e}_{i}\right),\\
 & \dot{\mathbf{B}}_{D}=0.
\end{aligned}
\label{eq:dSysHv}
\end{equation}
where 
$\mathbf{e}_{i}$ is the unit vector in the $i$-th Cartesian
direction.  It is easy to know the exact update of this subsystem  with
step size $\Delta t$ is
\begin{equation}
\phi^{v_i}(\Delta t):\begin{aligned} & \mathbf{X}_{s}(t+\Delta t)=\mathbf{X}_{s}(t)+V_{s}[i](t)\mathbf{e}_{i},\\
 & \mathbf{V}_{s}(t+\Delta t)=\mathbf{V}_{s}(t)+\mathbf{e}_{i}\times\left(\mathcal{W}_{b}^{S}(\mathbf{X}_{s})\mathbf{B}_{D}\right),\\
 & \mathbf{E}_{D}(t+\Delta t)=\mathbf{E}_{D}(t)-\mathcal{W}_{e}^{-1}\left(\sum_{s}V_{s}[i]\omega_{s}F_{J}(\mathbf{X}_{s},\Delta t)\mathbf{e}_{i}\right),\\
 & \mathbf{B}_{D}(t+\Delta t)=\mathbf{B}_{D}(t).\\
 & F(\mathbf{X}_{s},\Delta t)=\int_{X_{s}[i](t)}^{X_{s}[i](t+\Delta t)}\mathcal{W}_{e}^{S}(\mathbf{X}_{s})dX_{s}[i].
\end{aligned}
\label{eq:dSolHv}
\end{equation}
These exact solutions can be composed to get integrators for the semi-discrete system (\ref{eq:semiVM}). 
For example, a first order method can be constructed by
\[
\Phi(\Delta t)=\phi^{E}(\Delta t)\circ\phi^{B}(\Delta t)\circ\phi^{v1}(\Delta t)\circ\phi^{v2}(\Delta t)\circ\phi^{v3}(\Delta t),
\]
and a second order symmetric method can be derived from
\begin{align*}
\Phi(\Delta t)= & \phi^{E}(\Delta t/2)\circ\phi^{B}(\Delta t/2)\circ\phi^{v1}(\Delta t/2)\circ\phi^{v2}(\Delta t/2)\circ\phi^{v3}(\Delta t)\\
 & \circ\phi^{v2}(\Delta t/2)\circ\phi^{v1}(\Delta t/2)\circ\phi^{B}(\Delta t/2)\circ\phi^{E}(\Delta t/2).
\end{align*}
Higher order methods can be constructed by various ways of compositions\cite{Mclachlan02}. From the above expressing, it is known that the kind of methods can be computed explicitly,  hence are  easy to be implemented.

It can be verified that all the subsystems  satisfy $$\nabla\cdot\left(\sumli_{j}B_j(t)\mathbf{W}_j^b\right)=\nabla\cdot\left(\sumli_{j}B_j(0)\mathbf{W}_j^b\right),$$  and share the same bracket as the system (\ref{eq:semiVM}). According to the theory of Lie groups\cite{Olver93-242}, if the initial conditions are chosen as in Eq.~(\ref{eq:initial}), the numerical magnetic fields are divergence-free, and the discrete non-canonical Poisson structure is preserved by the methods. 

\section{Conclusion}
Based  on the Poisson bracket, we have developed Hamiltonian Particle-in-cell methods for Vlasov-Maxwell equations by combining the PIC technique with conforming finite element methods in space.
In order that the semi-discrete system conserves the discrete Poisson structure,
the finite element space  needs to satisfy
$\nabla\times \mathcal{E}_h\subset \mathcal{B}_h$, and $\nabla\cdot \mathbf{B}_h=0$ with $\mathcal{E}_h$ and $\mathcal{B}_h$ the  element spaces for the electric field and the magnetic field, respectively. Then finite elements for Maxwell's equations such as Raviart--Thomas--N\'{e}d\'{e}lec's mixed elements can be applied. The numerical methods which can preserve  the discrete Poisson bracket are constructed for the semi-discrete system by a Hamiltonian splitting method. We have given a general procedure of constructing Hamiltonian methods for the VM equations by FEM.  Numerical applications and error and stability analysis of the methods will be reported in future publications.

\begin{acknowledgements}
This research was supported by ITER-China Program
(2015GB111003, 2014GB124005, and 2013GB111000),
JSPS-NRF-NSFC A3 Foresight Program in the field of
Plasma Physics (NSFC-11261140328), the National Science
Foundation of China (11271357, 11575186, 11575185, 11505185, and
11505186), the CAS Program for Interdisciplinary
Collaboration Team, the Foundation for Innovative Research Groups of the NNSFC (11321061), the Geo-Algorithmic Plasma Simulator (GAPS) Project, and the U.S. Department of Energy (DEAC02-09CH11466).
\end{acknowledgements}

\appendix

\section{Derivation of the discrete Poisson bracket}\label{app:1}
It is known that the continuous VM equations is Hamiltonian characterized by the MMW Poisson bracket\cite{Marsden82,Morrison80},
\begin{equation}
\begin{aligned}\{\{\mathcal{F},\mathcal{G}\}\} & (f,\mathbf{E},\mathbf{B})=\int f\left\{ \frac{\delta\mathcal{F}}{\delta f},\frac{\delta\mathcal{G}}{\delta f}\right\} _{\mathbf{xv}}d\mathbf{x}d\mathbf{v}\\
 & +\int\left[\frac{\delta\mathcal{F}}{\delta\mathbf{E}}\cdot\left(\triangledown\times\frac{\delta\mathcal{G}}{\delta\mathbf{B}}\right)-\frac{\delta\mathcal{G}}{\delta\mathbf{E}}\cdot\left(\triangledown\times\frac{\delta\mathcal{F}}{\delta\mathbf{B}}\right)\right]d\mathbf{x}\\
 & +\int\left(\frac{\delta\mathcal{F}}{\delta\mathbf{E}}\cdot\frac{\partial f}{\partial\mathbf{v}}\frac{\delta\mathcal{G}}{\delta f}-\frac{\delta\mathcal{G}}{\delta\mathbf{E}}\cdot\frac{\partial f}{\partial\mathbf{v}}\frac{\delta\mathcal{F}}{\delta f}\right)d\mathbf{x}d\mathbf{v}\\
 & +\int f\mathbf{B}\cdot\left(\frac{\partial}{\partial\mathbf{v}}\frac{\delta\mathbf{\mathrm{\mathcal{F}}}}{\delta f}\times\frac{\partial}{\partial\mathbf{v}}\frac{\delta\mathcal{G}}{\delta f}\right)d\mathbf{x}d\mathbf{v},
\end{aligned}
\label{eq:MMWB}
\end{equation}
and the Hamiltonian
\begin{equation*}
\mathcal{H}(f,\mathbf{E},\mathbf{B})=\frac{1}{2}\int\mathbf{v}{}^{2}fd\mathbf{x}d\mathbf{v}+\frac{1}{2}\int\left(\mathbf{E}^{2}+\mathbf{B}^{2}\right)d\mathbf{x}.\label{eq:HamiltonVM}
\end{equation*}
Here the operator $\left\{ \cdot,\cdot\right\} _{\mathbf{xv}}$
denotes the canonical Poisson bracket for  functions of $\left(\mathbf{x},\mathbf{v}\right)$.
During spatial dicretisation, the variable $f$ is approximated by Eq.~(\ref{eq:Disf}), and $\mathbf{E}$ and $\mathbf{B}$ are approximated by Eq.~(\ref{eq:DisEB}). The Poisson bracket can be discretised in the same way. Firstly we deal with the dicretisation of the variation, by the approximate expression for the variables, and the chain rule of variations.
From Eq.~(\ref{eq:Disf}), the   discrete variables can be reexpressed by
$$\omega_{s}=\int f_{s}d\mathbf{x}d\mathbf{v},
\mathbf{X}_{s}=\frac{1}{\omega_{s}}\int\mathbf{x}f_{s}d\mathbf{x}d\mathbf{v},
\mathbf{V}_{s}=\frac{1}{\omega_{s}}\int\mathbf{v}f_{s}d\mathbf{x}d\mathbf{v}.$$
It follows that
\begin{gather*}
\frac{\delta\omega_{s}}{\delta f_{s}}=1,\quad\frac{\delta\mathbf{X}_{s}}{\delta f_{s}}=\frac{\mathbf{x}-\mathbf{X}_{s}}{\omega_{s}},\quad\frac{\delta\mathbf{V}_{s}}{\delta f_{s}}=\frac{\mathbf{v}-\mathbf{V}_{s}}{\omega_{s}}.
\end{gather*}
Using chain rule of variations leads to
\begin{equation}
\begin{aligned}\frac{\delta}{\delta f_{s}} & =\frac{\delta\omega_{s}}{\delta f_{s}}\frac{\partial}{\partial\omega_{s}}+\frac{\delta\mathbf{X}_{s}}{\delta f_{s}}\frac{\partial}{\partial\mathbf{X}_{s}}+\frac{\delta\mathbf{V}_{s}}{\delta f_{s}}\frac{\partial}{\partial\mathbf{V}_{s}}\\
 & =\frac{\partial}{\partial\omega_{s}}+\frac{\mathbf{x}-\mathbf{X}_{s}}{\omega_{s}}\frac{\partial}{\partial\mathbf{X}_{s}}+\frac{\mathbf{v-}\mathbf{V}_{s}}{\omega_{s}}\frac{\partial}{\partial\mathbf{V}_{s}}.
\end{aligned}
\label{eq:variationalfs}
\end{equation}
To be short, in the following we replace the notation $\mathbf{E}$ as for $\mathbf{E}_h$. 
From the discretisation for the fields in Eq.~(\ref{eq:DisEB}),
$$\left(\mathbf{E},\mathbf{W}_{j}^{e}\right)=\sum_{i=1}^{N_e}E_i\left(\mathbf{W}_{i}^{e},\mathbf{W}_{j}^{e}\right). $$
Denote $(\mathcal{W}_{e})_{ij}=\left(\mathbf{W}_{i}^{e},\mathbf{W}_{j}^{e}\right)$, 
then taking the variation of $E_{i}(t)$  w.r.t. $\mathbf{E}(\mathbf{x},t)$
gives
\[
\frac{\delta{E_{i}}}{\delta\mathbf{E}}=\sum\limits _{j}(\mathcal{W}_{e}^{-1})_{ij}\mathbf{W}_{j}^{e},\quad i=1\ldots N_{e}.
\]
It follows from the chain rule of variation that
\begin{equation}
\begin{aligned}\frac{\delta}{\delta\mathbf{E}}=\sum\limits _{i}\frac{\delta E_{i}}{\delta\mathbf{E}}\frac{\partial}{\partial E_{i}}=\sum_{i,j}\frac{\partial}{\partial E_{i}}(\mathcal{W}_{e}^{-1})_{ij}\mathbf{W}_{j}^{e}.
\end{aligned}
\label{eq:variationalE}
\end{equation}

 As follows,
we apply the above functional derivative to derive the discrete bracket corresponding to Eq.~(\ref{eq:MMWB}).

Consider the first term in Eq.~(\ref{eq:MMWB}). 
Substituting Eq.~(\ref{eq:variationalfs}) in the canonical bracket gives
\begin{align*}
\left\{ \frac{\delta\mathcal{F}}{\delta f_{s}},\frac{\delta\mathcal{G}}{\delta f_{s}}\right\} _{\bf xv} & =\left\{ \frac{\partial F}{\partial\omega_{s}}+\frac{\mathbf{x}-\mathbf{X}_{s}}{\omega_{s}}\frac{\partial F}{\partial\mathbf{X}_{s}}+\frac{\mathbf{v-}\mathbf{V}_{s}}{\omega_{s}}\frac{\partial F}{\partial\mathbf{V}_{s}},\frac{\partial G}{\partial\omega_{s}}+\frac{\mathbf{x}-\mathbf{X}_{s}}{\omega_{s}}\frac{\partial G}{\partial\mathbf{X}_{s}}+\frac{\mathbf{v-}\mathbf{V}_{s}}{\omega_{s}}\frac{\partial G}{\partial\mathbf{V}_{s}}\right\} _{\bf xv}\\
 & =\frac{1}{\omega_{s}^{2}}\left(\frac{\partial F}{\partial\mathbf{X}_{s}}\cdot\frac{\partial G}{\partial\mathbf{V}_{s}}-\frac{\partial F}{\partial\mathbf{V}_{s}}\cdot\frac{\partial G}{\partial\mathbf{X}_{s}}\right),
\end{align*}
so

$$\begin{aligned}
\sum_{s}\int f_{s}\left\{ \frac{\delta\mathcal{F}}{\delta f_{s}},\frac{\delta\mathcal{G}}{\delta f_{s}}\right\} _{\mathbf{xv}}d\mathbf{x}d\mathbf{v}= & \sum_{s}\frac{1}{\omega_{s}}\left(\frac{\partial F}{\partial\mathbf{X}_{s}}\cdot\frac{\partial G}{\partial\mathbf{V}_{s}}-\frac{\partial F}{\partial\mathbf{V}_{s}}\cdot\frac{\partial G}{\partial\mathbf{X}_{s}}\right).
\end{aligned}\eqno(T1)$$

Next we consider the second term of Eq.~(\ref{eq:MMWB}). Using Eq.~(\ref{eq:variationalE}) and the fact that $\mathcal{W}_{b}^{-1}$ is symmetric, we have
\begin{align*}
  \int\nabla\times \left(\frac{\delta{E_{i}}}{\delta\mathbf{E}}\right)\cdot\frac{\delta{B_{j}}}{\delta\mathbf{B}} d\mathbf{x}
  =&\sum\limits _{m,l}(\mathcal{W}_{e}^{-1})_{im}\int\nabla\times\mathbf{W}_{m}^{e}\cdot \mathbf{W}_{l}^{b} d\mathbf{x}(\mathcal{W}_{b}^{-1})_{jl}\\
  =&\sum\limits _{l,m}(\mathcal{W}_{e}^{-1})_{im}\mathcal{K}_{ml}(\mathcal{W}_{b}^{-1})_{lj},
\end{align*}
where the notation $\mathcal{K}_{ml}=\int\nabla\times\mathbf{W}_{m}^{e}\cdot \mathbf{W}_{l}^{b} d\mathbf{x}$. So in the second term there is
$$\begin{aligned}
  \int\frac{\delta\mathcal{F}}{\delta\mathbf{E}}\cdot\left(\nabla\times\frac{\delta\mathcal{G}}{\delta\mathbf{B}}\right)d\mathbf{x}
=&\int\left(\nabla\times\frac{\delta\mathcal{F}}{\delta\mathbf{E}}\right)\cdot\frac{\delta\mathcal{G}}{\delta\mathbf{B}}d\mathbf{x}\\
 =&\sumli_{i,j}\frac{\delta\mathcal{F}}{\delta E_i}\frac{\delta\mathcal{G}}{\delta B_j}\int \nabla\times \left(\frac{\delta{E_{i}}}{\delta\mathbf{E}}\right)\cdot\frac{\delta{B_{j}}}{\delta\mathbf{B}} d\mathbf{x}\\
= & \left(\frac{\partial F}{\partial\mathbf{E}_{D}}\right)^{\mathrm{T}}\mathcal{W}_{e}^{-1}\mathcal{K}\mathcal{W}_{b}^{-1}\frac{\partial G}{\partial\mathbf{B}_{D}}.
\end{aligned}\eqno(T2)$$

When the third term in Eq.~(\ref{eq:MMWB}) is considered, with the boundary condition  $f|_{\partial \Omega}=0$, we can use  the integration by part and to calculate the following term
\[
\int f_{s}\left(\frac{\delta\mathcal{G}}{\delta\mathbf{E}}\cdot\frac{\partial}{\partial\mathbf{v}}\frac{\delta\mathcal{F}}{\delta f_{s}}-\frac{\delta\mathcal{F}}{\delta\mathbf{E}}\cdot\frac{\partial}{\partial\mathbf{v}}\frac{\delta\mathcal{G}}{\delta f_{s}}\right)d\mathbf{x}d\mathbf{v}.
\]
It is noticed  by using Eq.~(\ref{eq:variationalfs}) that
\[
\frac{\partial}{\partial\mathbf{v}}\frac{\delta\mathcal{G}}{\delta f_{s}}=\frac{1}{\omega_{s}}\frac{\partial G}{\partial\mathbf{V}_{s}}.
\]
Then substitute the above relation and Eq.~(\ref{eq:variationalE}) in the third term, we have
$$\begin{aligned}
\int f_{s}\left(\frac{\delta\mathcal{F}}{\delta\mathbf{E}}\cdot\frac{\partial}{\partial\mathbf{v}}\frac{\delta\mathcal{G}}{\delta f_{s}}\right)d\mathbf{x}d\mathbf{v} &
=\frac{1}{\omega_{s}}\int f_{s}\left(\sum_{i,j}\frac{\partial F}{\partial E_i}\left(\mathcal{W}_{e}^{-1}\right)_{ij}\mathbf{W}_j^{e}(\mathbf{x})\cdot\frac{\partial G}{\partial\mathbf{V}_{s}}\right)d\mathbf{x}d\mathbf{v}\\
&=\sum_{i,j}\frac{\partial F}{\partial E_i}\left(\mathcal{W}_{e}^{-1}\right)_{ij} \left(\int \mathbf{W}_j^{e}(\mathbf{x})\delta(\mathbf{x}-\mathbf{X}_s)d\mathbf{x}d\mathbf{v}\right)\cdot\frac{\partial G}{\partial\mathbf{V}_{s}}\\
 & =\left(\frac{\partial F}{\partial\mathbf{E}_{D}}\right)^{\mathrm{T}}\mathcal{W}_{e}^{-1}\mathcal{W}_{e}^{S}(\mathbf{X}_{s})\frac{\partial G}{\partial\mathbf{V}_{s}},
\end{aligned}\eqno(T3)$$
where $\mathcal{W}_{e}^{S}(\mathbf{X}_{s})$ is the $N_{e}\times3$ matrix
function with the $j$-th row being $\int_{\Omega_{x}}\left(\mathbf{W}_{j}^{e}\right)^{\top}\delta(\mathbf{x}-\mathbf{X}_{s})d\mathbf{x}$.

For the fourth term of Eq.~(\ref{eq:MMWB}), the substitution of the expressions in Eqs.~(\ref{eq:DisEB}), (\ref{eq:variationalfs}), and (\ref{eq:variationalE}) derives
$$\begin{aligned}
\int f_{s}\mathbf{B}\cdot\left(\frac{\partial}{\partial\mathbf{v}}\frac{\delta\mathbf{\mathrm{\mathcal{F}}}}{\delta f_{s}}\times\frac{\partial}{\partial\mathbf{v}}\frac{\delta\mathcal{G}}{\delta f_{s}}\right)d\mathbf{x}d\mathbf{v}
& =\int f_{s}\sum_jB_j\mathbf{{W}}_j^b(\mathbf{x})\cdot\left(\frac{1}{\omega_{s}^{2}}\frac{\partial F}{\partial\mathbf{V}_{s}}\times\frac{\partial G}{\partial\mathbf{V}_{s}}\right)d\mathbf{x}d\mathbf{v}\\
 & =\frac{1}{\omega_{s}}\sum_j{B}_{j}\int\mathbf{{W}}_j^{b}(\mathbf{x})\delta(\mathbf{x}-\mathbf{X}_{s})d\mathbf{x}\cdot\left(\frac{\partial F}{\partial\mathbf{V}_{s}}\times\frac{\partial G}{\partial\mathbf{V}_{s}}\right)\\
 & =\frac{1}{\omega_{s}}\mathbf{B}_{D}^{T}\mathcal{W}_{b}^{S}(\mathbf{X}_{s})\left(\frac{\partial F}{\partial\mathbf{V}_{s}}\times\frac{\partial G}{\partial\mathbf{V}_{s}}\right),
\end{aligned}\eqno(T4)$$
where $\mathcal{W}_{b}^{S}(\mathbf{X}_{s})$ is a $N_{b}\times3$ matrix
function with the $j$-th row being $\int_{\Omega_{x}}\left(\mathbf{W}_{j}^{b}\right)^{\top}\delta(\mathbf{x}-\mathbf{X}_{s})d\mathbf{x}$.
The summation of the terms (T1)--(T4) forms the discrete Poisson bracket in Eq.~(\ref{eq:dPoisson}).
\section{Proof of Lemma.\ref{th:Poisson}}\label{app:2}
A bracket operator is Poisson if it is bilinear, skew-symmetric ($\{F,G\}=-\{G,F\}$), and satisfies the Jacobi identity
$$\left\{ \left\{ F,G\right\} ,H\right\} +\left\{ \left\{ G,H\right\} ,F\right\}  +\left\{ \left\{ H,F\right\} ,G\right\}  =0.$$
The skew-symmetry of the bracket defined in Eq.~(\ref{eq:dPoisson}) is easy to be verified. Next we prove the Jacobi-identity of the bracket.

Recall that any finite dimensional Poisson bracket can be represented and characterized by a matrix. We rewrite the bracket in Eq.~(\ref{eq:dPoisson}) as four parts,
\[
\left\{ F,G\right\} =\left\{ F,G\right\} _{xv}+\left\{ F,G\right\} _{EB}+\left\{ F,G\right\} _{Ev}+\left\{ F,G\right\} _{B},
\]
each part is in the form
\begin{align*}
 & \left\{ F,G\right\} _{xv}=\sum_{s}\left(\frac{\partial F}{\partial(\mathbf{X}_{s},\mathbf{V}_{s})}\right)^{T}\mathcal{J}\left(\frac{\partial G}{\partial(\mathbf{X}_{s},\mathbf{V}_{s})}\right),\\
 & \left\{ F,G\right\} _{EB}=\left(\frac{\partial F}{\partial(\mathbf{{E}}_{D},\mathbf{B}_{D})}\right)^{T}\mathcal{L}_{EB}\left(\frac{\partial G}{\partial(\mathbf{E}_{D},\mathbf{B}_{D})}\right),\\
 & \left\{ F,G\right\} _{Ev}=\sum_{s}\left(\frac{\partial F}{\partial(\mathbf{V}_{s},\mathbf{E}_{D})}\right)^{T}\mathcal{\mathcal{L}}_{Ev}(\mathbf{X}_{s})\left(\frac{\partial G}{\partial(\mathbf{V}_{s},\mathbf{E}_{D})}\right),\\
 & \left\{ F,G\right\} _{B}=\sum_{s}\frac{1}{\omega_{s}}\mathbf{B}_{D}^{T}\mathcal{W}_{b}^{S}(\mathbf{X}_{s})\left(\frac{\partial F}{\partial\mathbf{V}_{s}}\times\frac{\partial G}{\partial\mathbf{V}_{s}}\right),
\end{align*}
with the matrices being
\begin{subequations}
\begin{gather}
\mathcal{J}=\frac{1}{\omega_{s}}\left(\begin{array}{cc}
0 & I_{3}\\
-I_{3} & 0
\end{array}\right),\\
\mathcal{L}_{EB}=\left(\begin{array}{cc}
0 & \mathcal{W}_{e}^{-1}\mathcal{K}\mathcal{W}_{b}^{-1}\\
-\mathcal{W}_{b}^{-1}\mathcal{K^{\mathrm{T}}}\mathcal{W}_{e}^{-1} & 0
\end{array}\right),\\
\mathcal{L}_{Ev}(\mathbf{X}_{s})=\left(\begin{array}{cc}
0 & \mathcal{W}_{e}^{S}(\mathbf{X}_{s})^{T}\mathcal{W}_{e}^{-1}\\
-\mathcal{W}_{e}^{-1}\mathcal{W}_{e}^{S}(\mathbf{X}_{s}) & 0
\end{array}\right).\\
\nonumber
\end{gather}
\end{subequations}
The properties of the matrices determines those of the bracket. Before complete the proof, we need the following lemma derived from ChapterVII,\cite{Hairer03}.
\begin{lemma}\label{lem:Hairer}
The Poisson bracket (\ref{eq:dPoisson}) satisfies the Jacobi-identity for all functions $F$, $G$, $K$ if and only if it satisfies the identity for the coordinate functions $X_s,V_s,E_j,B_j$.
\end{lemma}
According to Lemma.~\ref{lem:Hairer}, in $\{\{F,G\},H\}$ one need only to consider the explicit dependence of the matrices on the variables when taking the derivatives of $\{F,G\}$.
Following the similar idea in Ref.~\onlinecite{Morrison13}, the Jacobi identity reads
\begin{equation*}
\begin{aligned}\left\{ \left\{ F,G\right\} ,H\right\}+cyc=  & \underbrace{\left\{ \left\{ F,G\right\} _{xv},H\right\} _{xv}}_{1}+\underbrace{\left\{ \left\{ F,G\right\} _{xv},H\right\} _{B}}_{2}\\
 & +\underbrace{\left\{ \left\{ F,G\right\} _{xv},H\right\} _{Ev}}_{3}+\underbrace{\left\{ \left\{ F,G\right\} _{xv},H\right\} _{EB}}_{4}\\
 & +\underbrace{\left\{ \left\{ F,G\right\} _{B},H\right\} _{xv}}_{5}+\underbrace{\left\{ \left\{ F,G\right\} _{B},H\right\} _{B}}_{6}\\
 & +\underbrace{\left\{ \left\{ F,G\right\} _{B},H\right\} _{Ev}}_{7}+\underbrace{\left\{ \left\{ F,G\right\} _{B},H\right\} _{EB}}_{8}\\
 & +\underbrace{\left\{ \left\{ F,G\right\} _{Ev},H\right\} _{xv}}_{9}+\underbrace{\left\{ \left\{ F,G\right\} _{Ev},H\right\} _{B}}_{10}\\
 & +\underbrace{\left\{ \left\{ F,G\right\} _{Ev},H\right\} _{Ev}}_{11}+\underbrace{\left\{ \left\{ F,G\right\} _{Ev},H\right\} _{EB}}_{12}\\
 & +\underbrace{\left\{ \left\{ F,G\right\} _{EB},H\right\} _{xv}}_{13}+\underbrace{\left\{ \left\{ F,G\right\} _{EB},H\right\} _{B}}_{14}\\
 & +\underbrace{\left\{ \left\{ F,G\right\} _{EB},H\right\} _{Ev}}_{15}+\underbrace{\left\{ \left\{ F,G\right\} _{EB},H\right\} _{EB}}_{16}+cyc,
\end{aligned}
\end{equation*}
where the symbol `cyc' means cyclic permutation. By Lemma.~\ref{lem:Hairer}, apparently we have the following results:
\begin{itemize}
\item Term 1-4 vanishes as $\mathcal{J}$ in $\{F,G\}_{xv}$ is constant.
\item Term 6-7 vanishes as $\mathcal{W}_{b}^{S}(X_{s})$ in $\{F,G\}_{B}$
is independent of $E_{J}$ and $V_{s}$.
\item Term 10-12 vanishes as $\mathcal{L}_{Ev}$ in $\{F,G\}_{Ev}$ is independent
of $E_{J},B_{J}$ and $V_{s}$.
\item Term 13-16 vanishes as $\mathcal{L}_{EB}$ in $\{F,G\}_{EB}$ is constant.
\end{itemize}
Therefore, only the terms 5, 8 and 9 need to be analyzed.

Term 5 reads
\begin{equation}\label{eq:term5}
\begin{aligned}
\left\{ \left\{ F,G\right\} _{B},H\right\} _{xv}+cyc & =\sum_{s}\frac{1}{\omega_{s}}\frac{\partial\left\{ F,G\right\} _{B}}{\partial X_{s}}\cdot\frac{\partial H}{\partial V_{s}}+cyc\\
 & =\sum_{s=1}^{N}\frac{1}{\omega_{s}^{2}}\sum_{i=1}^{3}\left(\mathbf{B}_{D}^{T}\partial_{X_{s}^{i}}\mathcal{W}_{b}^{S}(\mathbf{X}_{s})\right) \left(\frac{\partial F}{\partial\mathbf{V}_{s}}\times\frac{\partial G}{\partial\mathbf{V}_{s}}\right)\frac{\partial H}{\partial V_{s}^{i}}+cyc\\
 & =\sum_{s=1}^{N}\frac{1}{\omega_{s}^{2}}\nabla_{X_{s}}\cdot\left(\mathbf{B}_{D}^{T}\mathcal{W}_{b}^{S}(\mathbf{X}_{s})\right)\left(\frac{\partial F}{\partial\mathbf{V}_{s}}\times\frac{\partial G}{\partial\mathbf{V}_{s}}\right)\cdot\frac{\partial H}{\partial V_{s}},
\end{aligned}
\end{equation}
by the property of cross production, $\left(a\times b\right)_{i}c_{j}+cyc=\delta_{ij}(a\times b)\cdot c$.
The above Jacobian relation vanishes for
\[
\nabla_{X_{s}}\cdot\left(\mathbf{B}_{D}^{T}\mathcal{W}_{b}^{S}(\mathbf{X}_{s})\right)=0.
\]
Applying  the  integration by parts to the above equality  provides
\begin{align*}
0=\nabla_{X_{s}}\cdot\left(\mathbf{B}_{D}^{T}\mathcal{W}_{b}^{S}(\mathbf{X}_{s})\right)
&=\sum_{j=1}^{N_{b}}B_{j}\nabla_{X_{s}}\cdot \int\mathbf{W}^b_{j}\delta(\mathbf{x}-\mathbf{X_s}) d{\bf x}\\
&=\sum_{j=1}^{N_{b}}B_{j}\int\nabla\cdot\mathbf{W}^b_{j}\delta(\mathbf{x}-\mathbf{X_s}) d{\bf x}.
\end{align*}

Now, we consider the terms 8 and 9. Term 9 is
\begin{equation}\label{eq:term9}
\begin{aligned}
 & \left\{ \left\{ F,G\right\} _{Ev},H\right\} _{xv}+cyc\\
= & \sum_{s}\frac{1}{\omega_{s}}\frac{\partial\left\{ F,G\right\} _{Ev}}{\partial X_{s}}\cdot\frac{\partial H}{\partial V_{s}}+cyc\\
= & \sum_{s}\frac{1}{\omega_{s}}\sum_{i=1}^{3}\left[\left(\frac{\partial G}{\partial\mathbf{E}_{D}}\right)^{T}\mathcal{W}_{e}^{-1}\partial_{X_{s}^{i}}\mathcal{W}_{e}^{S}\frac{\partial F}{\partial\mathbf{V}_{s}}\right]\frac{\partial H}{\partial V_{s}^{i}}-\left[\left(\frac{\partial F}{\partial\mathbf{E}_{D}}\right)^{T}\mathcal{W}_{e}^{-1}\partial_{X_{s}^{i}}\mathcal{W}_{e}^{S}\frac{\partial G}{\partial\mathbf{V}_{s}}\right]\frac{\partial H}{\partial V_{s}^{i}}+cyc\\
= & \sum_{s}\frac{1}{\omega_{s}}\left(\frac{\partial H}{\partial\mathbf{E}_{D}}\right)^{T}\mathcal{W}_{e}^{-1}M(\mathbf{X}_s)\left(\frac{\partial F}{\partial\mathbf{V}_{s}}\times\frac{\partial G}{\partial\mathbf{V}_{s}}\right)+cyc,
\end{aligned}
\end{equation}
where  $M(\mathbf{X}_s)$ denotes the $N_{e}\times3$ dimensional matrix with the $j$-th row being $\int(\nabla\times\mathbf{W}^e_{i})\delta(\mathbf{x}-\mathbf{X}_{s}) d\mathbf{x}$.
The final equality is derived by summarizing the permuted terms, and taking the integration by parts
\[
\nabla_{X_{s}}\times\int\mathbf{W}^e_{i}(\mathbf{x})\delta(\mathbf{x}-\mathbf{X}_{s})d\mathbf{x} 
=\int(\nabla\times\mathbf{W}^e_{i})^\top\delta d\mathbf{x},\quad\forall i,s.
\]

 Term 8 can be  written as
\begin{equation}\label{eq:term8}
\begin{aligned}
\left\{ \left\{ F,G\right\} _{B},H\right\} _{EB}+cyc & =-\left(\frac{\partial H}{\partial\mathbf{E}_{D}}\right)^{T}\mathcal{W}_{e}^{-1}\mathcal{K}\mathcal{W}_{b}^{-1}\frac{\partial\left\{ F,G\right\} _{B}}{\partial\mathbf{B}_{D}}+cyc\\
 & =-\sum_{s}\frac{1}{\omega_{s}}\left(\frac{\partial H}{\partial\mathbf{E}_{D}}\right)^{T}\mathcal{W}_{e}^{-1}\mathcal{K}\mathcal{W}_{b}^{-1}\mathcal{W}_{b}^{S}(\mathbf{X}_{s})\left(\frac{\partial F}{\partial\mathbf{V}_{s}}\times\frac{\partial G}{\partial\mathbf{V}_{s}}\right)+cyc.\\
\end{aligned}
\end{equation}
Thus the summation of Term 8 in Eq.~(\ref{eq:term8}) and Term 9 in Eq.~(\ref{eq:term9}) gives
\begin{align*}
 & \left\{ \left\{ F,G\right\} _{Ev},H\right\} _{xv}+\left\{ \left\{ F,G\right\} _{B},H\right\} _{EB}+cyc\\
= & \sum_{s}\frac{1}{\omega_{s}}\left(\frac{\partial H}{\partial\mathbf{E}}\right)^{T}\mathcal{W}_{e}^{-1}\left(M(\mathbf{X}_s)-\mathcal{K}\mathcal{W}_{b}^{-1}\mathcal{W}_{b}^{S}\right)\left(\frac{\partial F}{\partial\mathbf{V}_{s}}\times\frac{\partial G}{\partial\mathbf{V}_{s}}\right)+cyc.
\end{align*}
It vanishes when $M(\mathbf{X}_s)=\mathcal{K}\mathcal{W}_{b}^{-1}\mathcal{W}_{b}^{S}$,
which can be simplified as
\[
\int(\nabla\times\mathbf{W}^e_{i})^{\top}\delta d\mathbf{x}=\sum_{j=1}^{Nb}\left(\mathcal{K}\mathcal{W}_{b}^{-1}\right)_{ij}\int(\mathbf{W}^b_{j})^{\top}\delta d\mathbf{x},\quad\forall i,s.
\]

Overall, considering the conditions in term 5 and term 8+9 respectively, the discrete bracket is Poisson if the following two conditions holds,
\begin{subequations}
\begin{gather}
\int(\nabla\times\mathbf{W}^e_{i}) \delta d\mathbf{x}=\sum_{j=1}^{Nb}\left(\mathcal{K}\mathcal{W}_{b}^{-1}\right)_{ij}\int(\mathbf{W}^b_{j}) \delta d\mathbf{x},\quad\forall i,s\label{eq:PoissCondI2}\\
\sum_{j=1}^{N_{b}}B_{j}\int\nabla\cdot\mathbf{W}^b_{j}\delta d{\bf x}=0.\label{eq:PoissCondII2}
\end{gather}
\end{subequations}

 \bibliographystyle{apsrev4-1}
\bibliography{Refs}

\end{document}